\documentclass[11pt,twoside]{article}
\usepackage{asp2004}
\usepackage{psfig}
\usepackage{epsf}
\usepackage{graphics}
\usepackage{lscape}
\def\edcomment#1{\iffalse\marginpar{\raggedright\sl#1\/}\else\relax\fi}
\reversemarginpar

\begin{document}

\title{Mass, Age, and Space Distributions of Star Clusters}

\author{S. Michael Fall}

\affil{Space Telescope Science Institute, 3700 San Martin Drive,
Baltimore, MD, 21218, USA}

\begin{abstract}

This article reviews some recent studies of the mass, age,
and space distributions of star clusters, with a primary
focus on the large population of clusters in the interacting 
Antennae galaxies. 
Some of the highlights of these studies are the following:
1. The mass function of the young clusters (with ages $\tau 
< 10^8$~yr) has power-law form, $dN/dM \propto M^{-2}$, over 
the observed range of masses, $10^4 < M < 10^6$~$M_{\odot}$.
2. The age distribution of the clusters declines steeply at
all ages, roughly as $dN/d\tau \propto \tau^{-1}$ for 
mass-limited samples, indicating rapid disruption of most
clusters. 
3. At least 20\% and possibly all stars form in clusters 
and/or associations, including those that are unbound and 
short-lived.
4. Many of the clusters that remain bound just after their 
formation are disrupted on longer timescales by a combination
of mass loss by stellar evolution and several stellar 
dynamical processes. 
5. The young clusters have a clumpy space distribution and
are located preferentially in regions of high interstellar 
density, at least when averaged over scales of about a kpc.
6. The positions of the young clusters, however, are not
correlated with the local velocity gradients or velocity 
dispersions in the interstellar medium.
There are growing indications that most of these conclusions 
also apply to the populations of young star clusters in other 
galaxies, both interacting and quiescent. 

\end{abstract}
\thispagestyle{plain}

\section{Introduction}

Some of the most basic properties of star clusters are their
masses, ages, and positions. 
The distributions of these quantities for a population of 
clusters reveal clues about the processes involved in their
formation and disruption.
The focus of this article is on the mass, age, and space distributions
of the star clusters in the Antennae galaxies.
These clusters have attracted attention for several reasons.
The Antennae galaxies are the nearest and best-studied pair of 
merging galaxies, consisting of two large spirals that collided 
and began to commingle a ${\rm few} \times 10^8$~yr ago.
The number of young clusters in the Antennae galaxies is huge, 
permitting the mass, age, and space distributions to be determined 
better than in any other galaxies, except perhaps the Milky Way 
and Andromeda.
The ongoing merger is almost certainly responsible in some way 
for this large population of clusters.
Understanding the formation and disruption of clusters in this
setting is important because it represents a latter-day example
of the hierarchical formation of galaxies, a process that 
operated even more effectively in the early universe.
Thus, despite the apparently special nature of the cluster population
in the Antennae galaxies, the lessons we learn from it are likely to
have broad implications.

This article is a review of some studies made over the past five
years by Rupali Chandar, Fran{\c c}ois Schweizer, Bradley Whitmore, 
Qing Zhang, and the author (see also the articles in this volume
by Chandar and Whitmore).
The observational parts of these studies are based on images taken
in the broad $U$, $B$, $V$, $I$ and narrow H$\alpha$ passbands with
the Wide Field Planetary Camera 2 (WFPC2) on the {\it Hubble Space 
Telescope (HST)} and described fully by Whitmore et al. (1999).
We compare these observations with stellar population models to 
estimate the ages ($\tau$), extinctions ($A$), corrected 
luminosities ($L$), and hence masses ($M$) of the clusters. 
Some of the brightest clusters are spatially resolved in the 
{\it HST} images, but those near the limiting magnitude are 
indistinguishable from stars.
We minimize stellar contamination in our cluster sample by 
restricting it to objects brighter than the most luminous stars. 
Since our sample is optically selected, it undoubtly excludes 
the clusters most heavily obscured by dust. 
The fraction of missing clusters is hard to estimate but 
is expected to decline with increasing age.

\section{Mass Distribution}

Figure~1 shows the luminosity-age distribution of the star
clusters in the Antennae galaxies. This is an updated version 
of a similar diagram based on two reddening-free $Q$ parameters 
derived from our broad-band $U$, $B$, $V$, and $I$ measurements
(Zhang \& Fall 1999). The new diagram utilizes the narrow-band 
H$\alpha$ measurements, as well as the broad-band measurements,
and provides excellent age discrimination near $\tau \sim 10^7$~yr, 
the age at which the ionizing flux from a stellar population 
declines rapidly.
The vertical striations of the data points in Figure~1 are
artifacts caused by a combination of observational errors and
abrupt bends in the stellar population tracks in the color 
space in which the fits are made.
These small-scale features in the $L$-$\tau$ diagram should be 
ignored; only the gross distribution of data points is significant.
The diagonal lines in Figure~1 represent the evolutionary tracks 
of stellar population models of fixed initial mass, 
$M = 3\times10^4$~$M_{\odot}$ and $2\times10^5$~$M_{\odot}$. 
The horizontal line is close to the maximum stellar luminosity.

\begin{figure}[!ht]
\centerline{\psfig{figure=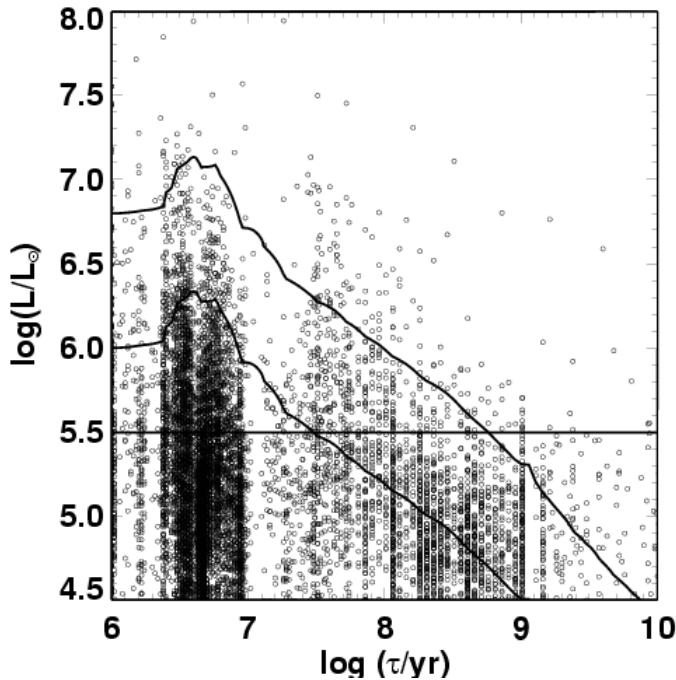,width=3.7in}}
\caption{Luminosity-age distribution of star clusters in the
Antennae galaxies (from Fall et al. 2004). 
$L$ is the extinction-corrected luminosity in the $V$-band. 
The diagonal lines are evolutionary tracks of stellar population
models with initial masses of $3\times10^4$~$M_{\odot}$ and
$2\times10^5$~$M_{\odot}$ (Bruzual \& Charlot 2003). 
The horizontal line at $L = 3\times10^5$~$L_{\odot}$ is the 
approximate upper limit for stellar contamination.
The narrow vertical features in the $L$-$\tau$ distribution
(stripes and gaps) are artifacts of the age-fitting procedure
(see text).}
\end{figure}

Figure~1 contains much of the available statistical information 
about the population of star clusters in the Antennae galaxies.
For example, the luminosity function, $\phi(L) \equiv dN/dL$, is
obtained by projecting the two-dimensional distribution horizontally
along the age axis.
We obtain the mass function, $\psi(M) \equiv dN/dM$, by projecting 
instead in a diagonal direction, along the stellar population tracks, 
and counting clusters in the corresponding mass bins.
It should be clear from this that the luminosity and mass functions
need not have the same or even similar forms.
The mass function, while more difficult to determine than the 
luminosity function (because the ages of the clusters must also
be estimated), is the more physically informative of the two.
The relationship between the luminosity and mass functions also
involves the age distribution, which in turn depends on the
formation and disruption histories of the clusters.
	
Figure~2 shows the mass function of the star clusters in the 
Antennae galaxies for two intervals of age, $10^6 < \tau < 10^7$~yr
and $10^7 < \tau < 10^8$~yr.
It is difficult to estimate $\psi(M)$ for older clusters because
only the most massive among them are brighter than the limiting
luminosity for stellar contamination (assumed for simplicity to 
be constant).
In Figure~2, we have plotted $\log \Psi$ against $\log M$, 
where $\Psi$, defined as the number of clusters per unit $\log M$, 
is related to $\psi$, the number of clusters per unit $M$, by
$\Psi(\log M) = (\log e)^{-1} M \psi(M)$.
Evidently, the mass function declines monotonically with increasing 
mass. 
It can be represented by a power law of the form $\psi(M) \propto
M^{\beta}$, with $\beta \approx -2$ over the entire observed range,
$10^4 < M < 10^6$~$M_{\odot}$.
Moreover, the exponent $\beta$ appears to have little or no dependence
on age, at least for $\tau < 10^8$~yr. (These results confirm the
earlier ones from Zhang \& Fall 1999.)
 
\begin{figure}[!ht]
\plotfiddle{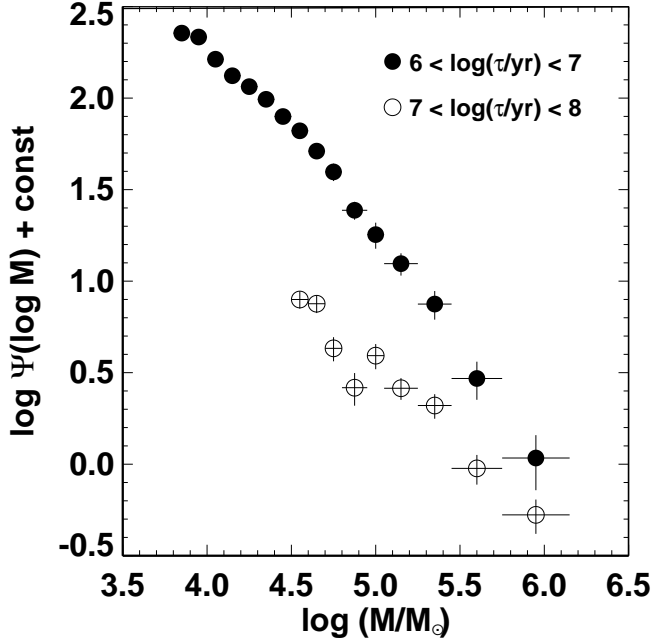}{3.7in}{0.}{50.}{50.}{-160.}{0}
\vspace{-0.5in}
\caption{Mass function of star clusters in the Antennae 
galaxies in two intervals of age (from Fall et al. 2004). 
The function is plotted, in each age interval, over the mass
range in which stellar contamination is negligible.
The best-fit power laws have exponents $\beta = -1.95 
\pm 0.03$ for $10^6 < \tau < 10^7$~yr and $\beta = -1.82 
\pm 0.08$ for $10^7 < \tau < 10^8$~yr.}
\end{figure}

The luminosity function of the star clusters in the Antennae galaxies
can also be represented, at least in a first approximation, by a
power law, $\phi(L) \propto L^{\alpha}$, with $\alpha \approx -2$ 
(Whitmore et al. 1999). 
The reason the mass and luminosity functions happen to have similar
forms in this case is that the age distribution is relatively narrow, 
most clusters being younger than $10^7$~yr (see below).
In the hypothetical limit that all clusters had exactly the same
age (and stellar initial mass function), $\phi(L)$ would have 
exactly the same dependence on $L$ that $\psi(M)$ has on $M$.

The power-law form of the mass function of the young clusters in 
the Antennae galaxies is reminiscent of the mass functions of the 
molecular clouds and complexes in the Milky Way (MW) and the Large 
Magellanic Cloud (LMC). 
These have $\psi(M) \propto M^{\beta}$, with $\beta \approx -2$ 
for $10^3 < M < 10^6$~$M_{\odot}$ in the MW (Heyer, Carpenter, 
\& Snell 2001) and $\beta \approx -2$ for $8 \times 10^4 < M < 
2 \times10^6$~$M_{\odot}$ in the LMC (Fukui 2002). 
Radio CO observations do not yet have enough angular resolution
to determine the mass functions of the molecular clouds in more
distant galaxies, such as the Antennae.
The power-law form of the mass functions of molecular clouds 
is almost certainly related in some way to the fractal or scale-free 
structure of the turbulent interstellar medium (ISM), perhaps 
the result of a density threshold (Elmegreen 2002). The exponent 
$\beta = -2$ has the property that equal logarithmic intervals 
of the masses of the clouds or clusters contain the same total mass.
Despite some recent progress, there does not yet appear to be
a compelling physical explanation for the particular value
$\beta \approx -2$.

The mass function of the young star clusters in the Antennae and
other galaxies is completely different from that of old globular 
clusters. For the latter, $\phi(L)$ and thus $\psi(M)$ are usually 
represented by lognormal distributions, with a peak at $M_p \approx 
2 \times 10^5$~$M_{\odot}$. It turns out, however, that the low-mass 
part ($M < M_p$) is fitted even better by $\psi(M) = {\rm const}$,
hence $\Psi(\log M) \propto M$ (McLaughlin 1994; Fall \& Zhang 2001).
Any population of star clusters will be eroded by a variety of 
disruptive processes, including mass loss by stellar evolution
and the evaporation of stars by internal two-body relaxation and 
external gravitational shocks. Of these, two-body relaxation is by 
far the dominant mechanism for low-mass clusters over long times
($\tau \ga {\rm few} \times 10^8$~yr). 
Several theoretical studies, spanning many years and with increasing 
realism, have shown that the disruption of clusters would cause 
their mass function, if initially a power law, to evolve in a 
Hubble time into one like that of old globular clusters (Fall 
\& Rees 1977; Baumgardt 1998; Vesperini 1998; Fall \& Zhang 2001).
This result provides a physical basis for the suggestion that
most or even all star clusters form by some universal mechanism
and that the differences in their observed properties mainly
reflect the differences in their ages and/or the sizes of the
samples (Elmegreen \& Efremov 1997; Larsen 2002; Whitmore 2003).

\section{Age Distribution}

The age distribution of a population of star clusters $dN/d\tau$
contains information about the formation and disruption of the
clusters. 
It can be derived from the two-dimensional $L$-$\tau$ diagram in 
different ways, depending on the selection criteria of the sample.
The most common form of $dN/d\tau$ is for a luminosity-limited 
sample, obtained simply by counting clusters in age bins above 
the limiting luminosity.
This, however, is not straightforward to interpret in dynamical
terms because it depends on the fading of the clusters by stellar
evolution in addition to their formation and disruption histories.
A more physically informative age distribution is that for a
mass-limited sample.
We obtain this form of $dN/d\tau$ by counting the clusters in age
bins above one of the stellar population tracks in Figure~1.

Figure~3 shows the age distribution of the star clusters in the 
Antennae galaxies for both luminosity- and mass-limited samples.
The former is somewhat steeper than the latter because it includes 
a higher proportion of young clusters and a lower proportion of
old clusters. 
In the following, we only consider results from mass-limited
samples, for the reasons mentioned above.
We find that the age distribution declines rapidly, starting at 
very young ages.
In particular, $dN/d\tau$ drops by a factor of at least 10
between $\tau \approx 10^6$~yr and $10^7$~yr.
The rate of decline has no noticeable dependence on mass, at least 
for $M > 3 \times 10^4$~$M_{\odot}$.
The age distribution has features corresponding to the vertical 
striations in Figure~1, which in turn are caused by a combination
of observational errors and abrupt bends in the stellar population 
tracks.
If we ignore these small-scale features and force-fit the age 
distribution by a power law, we obtain $dN/d\tau \propto \tau^{-1}$.
Including the clusters obscured by dust, presumably mostly
young, would steepen $dN/d\tau$.

\begin{figure}[!ht]
\plotfiddle{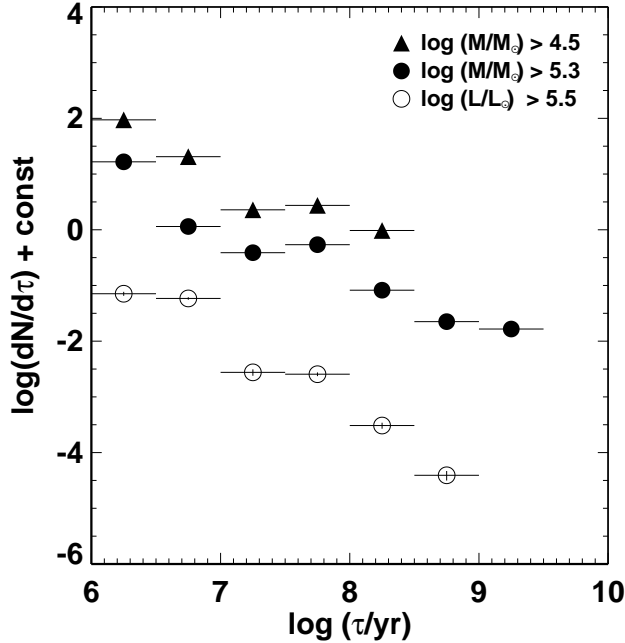}{3.7in}{0.}{50.}{50.}{-160.}{0}
\vspace{-0.5in}
\caption{Age distribution of star clusters in the Antennae
galaxies with different selection criteria (from Fall et al. 2004). 
The open symbols pertain to a luminosity-limited sample with
$L_V > 3 \times 10^5$~$L_{\odot}$, the approximate upper limit 
for stellar contamination, while the filled symbols pertain to
mass-limited samples with $M > 3\times10^4$~$M_{\odot}$ and
$M > 2\times10^5$~$M_{\odot}$.}
\end{figure}

The age distribution in general reflects a combination of birth 
and death rates.
Thus, the narrow peak near $\tau \sim 10^6$~yr might
be interpreted as the result of a very short and intense burst of
cluster formation, possibly triggered by the current interaction 
between the two Antennae galaxies. 
This, however, is extremely unlikely for two reasons.
First, the Antennae galaxies have been interacting for the past 
${\rm few} \times 10^8$~yr, a time much longer than the width of 
the peak in $dN/d\tau$. 
Second, the age distribution is similar in different parts of the
Antennae galaxies, some of which appear to be interacting more
strongly than others (see the article by Whitmore in this volume). 
Thus, it is almost certain that the steep decline in the age 
distribution is caused mainly by rapid disruption rather than 
a recent burst in the formation of the clusters. 

The short timescale on which the clusters are disrupted 
indicates that most of them are not gravitationally bound.
In terms of the initial characteristic radius $R_0$ 
(three-dimensional median radius) and virial velocity 
$V_0$, the crossing time for stars orbiting within a bound 
cluster or protocluster is $\tau_{\rm cr} = R_0/V_0$. 
We estimate $\tau_{\rm cr} \approx 10^6$~yr for a typical young 
cluster with $M \approx 10^5$~$M_{\odot}$, $R_0 \approx 5$~pc,
and $V_0 \approx (0.4GM/R_0)^{1/2} \approx 6$~kms$^{-1}$.
The crossing time is expected to be similar for clusters of 
different masses and radii because it depends on these quantities
only through the mean density, which is determined primarily by 
the tidal field of the host galaxy.
If a protocluster suddenly lost most of its mass by the removal 
of interstellar matter, it would no longer be gravitationally 
bound and would expand almost freely, its characteristic radius 
increasing with age as $R(\tau) \approx R_0 (\tau/\tau_{\rm cr})$ 
and its characteristic surface density decreasing as $\Sigma(\tau) 
\approx \Sigma_0(\tau/\tau_{\rm cr})^{-2}$.
Thus, after $\tau \sim 10 \tau_{\rm cr} \sim 10^7$~yr,
the surface brightness of the cluster (even ignoring the fading
by stellar evolution) would be roughly a factor of $10^2$ lower
or 5 magnitudes fainter than initially (at $\tau \sim 10^6$~yr), 
and it would then disappear among the statistical fluctuations 
in the foreground and background of field stars.
(Note that the half-light diameters of the youngest clusters 
are similar to the dimensions of a single WFC pixel, 0.1 arcsec; 
thus, the expressions above are expected to describe accurately
the apparent sizes and surface brightnesses of expanding 
clusters for $\tau \gg \tau_{\rm cr}$, but not for 
$\tau \sim \tau_{\rm cr}$.)

What could cause this high rate of infant mortality among
the star clusters in the Antennae and other galaxies?
The gravitational binding energy of a massive cluster 
is only $\sim$$10^{50}$~erg, much less than the energy output 
of a single massive star over its short lifetime. 
The energy and momentum output from massive stars comes in the 
form of ionizing radiation, stellar winds, jets, and supernovae. 
These processes could easily remove much of the ISM
from a protocluster, leaving the stars within it 
gravitationally unbound and expanding freely as argued above, 
even if the cloud in which they formed was initially bound.
The energy and momentum deposited in a protocluster by these 
processes is approximately proportional to the number of massive
stars and hence on average to the mass of the protocluster.
Moreover, the associated thermal and outflow velocities are 
much higher than the escape speed from the protoclusters. 
Thus, we expect the fraction of disrupted clusters to be roughly
independent of mass, consistent with our observations that the 
shape of the mass function is nearly independent of age, at
least for $\tau < 10^8$~yr, and that the shape of the age
distribution is nearly independent of mass, at least for
$M > 3 \times10^4$~$M_{\odot}$.

These arguments suggest that the survival of a cluster, not 
its disruption, may be the more difficult fact to explain.
Whether a particular cluster survives may depend on 
``accidental'' factors, such as just where and when the 
most massive stars happen to form within the protocluster.
Indeed, the inner, dense cores of protoclusters are more likely
to survive than their outer envelopes.
As a result, clusters may retain some of their stars and 
lose others.
Because the mass function of the clusters is a power law, roughly 
independent of age, we cannot distinguish between the case in which 
every cluster loses $\sim$90\% of its mass in the first $10^7$~yr
and the case in which $\sim$90\% of the clusters lose all of
their mass while the others lose none---or any other case between 
these extremes.
However, we can conclude, irrespective of this ambiguity, that
$\sim$90\% or more of the stars that form in recognizable 
clusters are dispersed in the field population
before they are $\sim$$10^7$~yr old. 
The few clusters that manage to survive their infancy are 
then subject to disruption on longer timescales by a
combination of stellar evolution and stellar dynamical 
processes, as discussed in Section~2.

In connection with the formation and disruption rates of the
clusters, it is also interesting to estimate the fraction of
stars that are born within clusters (and/or ``associations'').
We can set a lower limit on this fraction from the observed 
fraction of H$\alpha$ emission closely associated with the 
clusters in our sample.
This fraction is about 20\%.
This is a lower limit on the fraction of stars born in clusters
for several reasons: (1) it includes only those clusters brighter
than our stellar contamination limit and hence more massive than 
$\sim$$10^4$~$M_{\odot}$; (2) some of the ionizing radiation from 
the clusters will escape from their immediate vicinity and cause
H$\alpha$ emission elsewhere; (3) some clusters will be disrupted 
even before they stop producing ionizing radiation. 
Each of these effects alone could increase the fraction by factors
of 1--2, and together they could increase it by a factor up to 5.
We conclude from this that at least 20\% and possible all
of the stars in the Antennae galaxies were born in clusters,
including the large majority of clusters that were initially 
unbound.

\section{Space Distribution}

The space distribution of a population of star clusters, especially
when compared with the density and velocity fields of the nearby
interstellar medium, have the potential to reveal important clues 
about the physical processes involved in the the formation of the 
clusters.
We have made such a study for the star clusters in the Antennae 
galaxies (Zhang, Fall, \& Whitmore 2001).
This is based on the positions of the clusters in different age 
groups in our {\it HST} images and on the intensity and velocity maps 
of the galaxies at a variety of wavelengths from other investigations: 
broad-band radio (6~cm), infrared (60~$\mu$m and 15~$\mu$m), optical 
(8000~\AA), ultraviolet (1500~\AA), and X-ray (0.1--2.5~keV) continuum 
emission and narrow-band CO (1--0) 2.6~mm, HI 21~cm, and H$\alpha$ 
6563~\AA\ line emission.

We find that the clusters have a clumpy space distribution, 
with an angular autocorrelation function of approximate power-law 
form, $w(\theta) \propto \theta^{-1}$.
The cross-correlations between the positions of the clusters and
the other maps show, as expected, that the young clusters tend to 
be associated more with long-wavelength and less with short-wavelength
emission than the old clusters. 
Many of the youngest clusters appear to be embedded in molecular
cloud complexes with dimensions up to $\sim$1~kpc.
Both the surface rates of star and cluster formation,
$\Sigma_{\rm SFR}$ and $\Sigma_{\rm CFR}$, as determined by 
H$\alpha$ emission, are correlated with the surface density of 
the interstellar medium $\Sigma_{\rm ISM}$, the sum of the 
molecular and atomic densities, when averaged on scales of 
1~kpc or more.
These relations have much scatter, but in the mean, take the
approximate forms $\Sigma_{\rm SFR} \propto \Sigma_{\rm ISM}^{1.4}$
and $\Sigma_{\rm CFR} \propto \Sigma_{\rm ISM}^{1.4}$.
The first, for the star formation rate, is the usual Schmidt-Kennicutt 
relation; the second, for the cluster formation rate, is new.
The observation that $\Sigma_{\rm SFR}$ and $\Sigma_{\rm CFR}$ are
proportional to each other is, of course, consistent with our claim 
in the previous section that most or all stars form in clusters 
(including those that are unbound).
On scales smaller than 1~kpc, the scatter becomes so large that 
the correlations between $\Sigma_{\rm SFR}$, $\Sigma_{\rm CFR}$, 
and $\Sigma_{\rm ISM}$ practically disappear, possibly because the 
local ISM is depleted by recent star and cluster formation, possibly 
because the associated feedback by ionizing radiation, stellar winds, 
jets, and supernovae is also important.

We have also compared the positions of the star clusters in the
Antennae galaxies with the local velocity gradients and dispersions 
in the ISM, as derived from the HI 21~cm and H$\alpha$ 6563~\AA\ 
lines. 
This is motivated by suggestions that the main trigger for 
the formation of the clusters might be the 
motions in the ISM driven by the colliding galaxies.
The idea is that the the dynamic overpressure caused by these
motions might induce the collapse of clouds that would otherwise
be stable or long-lived in a more quiescent environment.
We find that some of the young clusters are located in regions 
with above-average velocity gradients and dispersions.
However, this is not statistically true of the population as 
a whole; the cross correlations between the positions of the 
clusters of all age groups and the interstellar velocity 
gradients and dispersions are nearly zero at all angular 
separations (see Figures 15, 16, and 17 of Zhang et al. 2001).
Evidently, the formation of the clusters depends more on the
density structure of the ISM than on its velocity structure. 

\section{Conclusions}

Some of the main conclusions of the studies reviewed here are 
listed below.
Strictly speaking, these pertain only to the star clusters in
the Antennae galaxies.
There is, however, growing evidence that many if not all of
these conclusions are more widely applicable and may even be
valid, at least approximately, for the majority star clusters 
in most galaxies (see, e.g., Larsen 2002; Lada \& Lada 2003;
Whitmore 2003, and references therein).

1. The mass function of the young clusters (with ages $\tau 
< 10^8$~yr) has power-law form, $dN/dM \propto M^{-2}$, over 
the observed range of masses, $10^4 < M < 10^6$~$M_{\odot}$. It 
is possible, even likely, that the mass function continues to 
rise toward lower masses. This is similar to the mass functions 
of the molecular clouds and young (open) star clusters in the
Milky Way and the Large Magellanic Cloud, but very different from
the mass functions of old (globular) clusters in the Milky Way
and other galaxies.

2. The age distribution of the clusters declines steeply at all 
ages, roughly as $dN/d\tau \propto \tau^{-1}$ for mass-limited 
samples, indicating rapid disruption of most clusters.
In particular, $\sim$90\% or more of the clusters disappear 
by an age of $\sim$$10^7$~yr, when the stars within them have 
completed fewer than $\sim$10 orbits. 
It is very likely that these clusters are not gravitationally
bound, and were disrupted near the times they formed by the 
energy and momentum input from young stars to the ISM
of the protoclusters. 

3. At least 20\% and possibly all stars form in 
clusters and/or associations, including those that are unbound 
and short-lived. However, because most clusters are disrupted 
while they are very young, the vast majority of stars spend 
most of their lives as members of the field population. 
As a result, bound, long-lived clusters contain only a small 
fraction of the stars in galaxies.

4. Many of the clusters that remain bound just after their
formation are disrupted on longer timescales by a combination
of mass loss by stellar evolution and several stellar 
dynamical processes. 
For low-mass clusters, the most important disruptive process 
is the evaporation of stars caused by internal, two-body 
relaxation.
Over a Hubble time, this can turn a mass function with an initial
power-law form into one like that of old globular clusters, with 
a peak at $M_p \approx 2 \times 10^5$~$M_{\odot}$. 

5. The young clusters have a clumpy space distribution and are
located preferentially in regions of high interstellar density, at 
least when averaged over scales of about a kpc. The correlation 
between the cluster formation rate and the ISM density is similar 
to that known previously between the star formation rate and the
ISM density, namely $\Sigma_{\rm CFR} \propto \Sigma_{\rm SFR} 
\propto \Sigma_{\rm ISM}^{1.4}$.

6. The positions of the young clusters, however, are not
correlated with the local velocity gradients or velocity dispersions 
in the interstellar medium. This somewhat surprising observation
places constraints on which physical mechanisms may have been
responsible for triggering the formation of the clusters. In
particular, the large-scale chaotic motions caused by the collision
and merging of the galaxies seems not to be the dominant mechanism. 

\acknowledgments{I am grateful to my collaborators on all the 
projects reviewed here, and especially to Rupali Chandar and 
Bradley Whitmore for help in the preparation of this article.}

\end{document}